**WILEY**

# Eternalism and the problem of hyperplanes

Matias Slavov

Philosophy, Faculty of Social Sciences, Tampere University, Tampere, Finland

**Correspondence**
Matias Slavov, Philosophy, Faculty of Social Sciences, Tampere University, P. O. Box FI-33014, Tampere, Finland.
Email: matias.slavov@tuni.fi

**Funding information**
Research funding provided by the Academy of Finland

**Abstract**

Eternalism is the view that the past, the present and the future exist *simpliciter*. A typical argument in favor of this view leans on the relativity of simultaneity. The 'equally real with' relation is assumed to be transitive between spacelike separated events connected by hyperplanes of simultaneity. This reasoning is in tension with the conventionality of simultaneity. Conventionality indicates that, even within a specific frame, simultaneity is based on the choice of the synchronization parameter. Hence the argument for eternalism is compromised. This paper lays out alternative eternalist strategies which do not hinge on hyperplanes. While we lack a rigorous proof for eternalism, there are still cogent reasons to prefer eternalism over presentism.

## 1 | INTRODUCTION

Traditional eternalist arguments apply the relativity of simultaneity together with hyperplanes that connect spacelike separated events. For these arguments to work, one needs to assume both ontological simultaneity and the transitivity of reality. Consider three distant events, *A*, *B* and *C*. *A* and *B* are in fact simultaneous. *B* is in fact simultaneous with *C*. *A* and *B* are co-real, so are *B* and *C*. Therefore, *A* and *C* are co-real. All events exist tenselessly in spacetime. Whether something is past, present or future is dependent upon the perspective. Someone's 'now' can be some other's past or future: the past, the present and the future are equally real.

The conventionality of simultaneity raises worrying issues for such arguments. If simultaneity is a matter of convention, different choices of the synchronization parameter yield different simultaneity relations. If hyperplanes are arbitrary constructions, arguments relying on ontological simultaneity and co-reality relations become questionable.

This article considers alternative strategies for the eternalist. To that end, the following structure is applied. Section 2 presents arguments that lean on hyperplanes. They provide rather straightforward reasons to subscribe to eternalism. Section 3 analyses the debate over the conventionality of simultaneity. If simultaneity is not ontological but instead based on our preference, the traditional eternalist arguments are compromised. Section 4 considers the conventionalist challenge. It shall be argued that there are good reasons to uphold eternalism even in the absence of hyperplanes. Section 4.1 clarifies the distinction between relativity and conventionality of simultaneity. The transitivity of reality is not questioned by conventionality. Relative quantities, like tensed locations, are both real and relative. Each observer has their own moment of 'now'. This reasoning is eternalist in spirit, as it treats tenses as perspectival and indexical. Section 4.2 makes the case for the distinction between appearance and reality by applying the relativity of simultaneity. Observers may perceive two events as simultaneous, but the actual physical events causing the observations could be successive (relative to an inertial frame). Although conventionality may not be completely thwarted, there are good reasons to think simultaneity is relative in the ontological sense. Accordingly, there is no unique present moment stretched throughout the entire universe. This suffices to render eternalism more credible than presentism.

## 2 | ARGUMENTS THAT DRAW ON HYPERPLANES

To apply Cornelis Willem Rietdijk's (1966, p. 341) notation, we may consider two inertial systems. $W_1$ at $O_1$ is co-real with $W_2$ at $O_2$ per $\{X_1, T_1\}$, while $W_2$ at $O_2$ is co-real with $W_1$ at $P_2$ per $\{X_2, T_2\}$. Therefore, $W_1$ at $O_1$ is equally real with their future version $W_1$ at $P_2$. Conversely, $W_1$ at $P_2$ is equally real with their past version $W_1$ at $O_1$ (Figure 1).

Applying hyperplanes to a relativistic spacetime diagram indicates, in a rather straightforward way, eternalism. The past, the present and the future all exist.

Less than six months after Rietdijk's paper, Hilary Putnam published an analogous result. To properly contextualize his argument, Putnam (1967, pp. 243–245) contrasts his view to Aristotle on future contingents. This is certainly a well-known topic which has spurred multiple commentaries throughout history (Øhrstrøm & Hasle, 2020). In the interpretation of Putnam, Aristotle was an indeterminist. The outcome of future events is not determined at the present time. Therefore, statements about the potential sea-fight tomorrow do not presently have a truth value. Jaakko Hintikka (1964, pp. 464–465) points out that sentences for Aristotle contain explicit or implicit references to the present time. They are not based on an objective chronology but depend on the moment of their utterance.

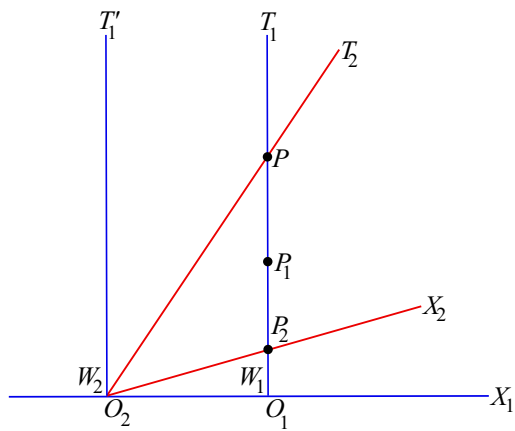

**FIGURE 1** A modification of Rietdijk's (1966, p. 341) diagram [Colour figure can be viewed at wileyonlinelibrary.com]

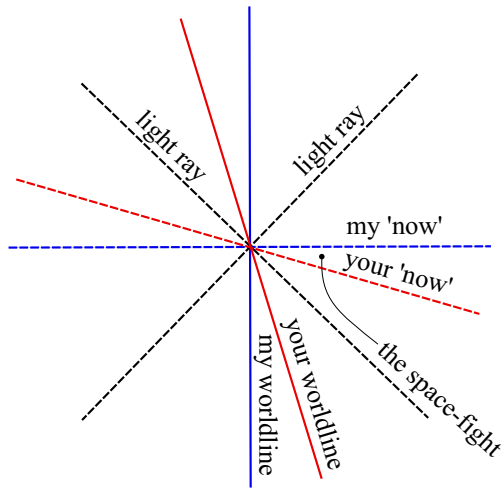

**FIGURE 2** Putnam's (1967, p. 241) diagram slightly modified [Colour figure can be viewed at wileyonlinelibrary.com]

For Aristotle the difference between the past and the future is that the former is determined whereas the latter indetermined. We do not know what the future will be but we may affect how it will unfold. We may know what the past is but we cannot affect how it lies. Putnam (1967, p. 244) claims 'Aristotle was wrong'. This is somewhat trivial as the special theory was formulated long after Aristotle. 'But it is important', Putnam (1967, p. 245) goes on, 'to see that Aristotle's view depends upon an *absolute* "pastness" and "futurity" just as much as Newtonian physics does, and that it is obsolete for the same reason'. To show this, Putnam refers to the following diagram (Figure 2).

Say there is a space-fight in 'your future' but in 'my past'. The fight is depicted as a point-event between our 'nows'.[1] If we adopt the Aristotelian view, in your coordinate system, the statement 'There will be a space-fight' has no truth value. However, in my coordinate system, the statement 'There was a space-fight' is true. Both statements concern the same physical event, the space-fight. There should be a truth on the matter of how the fight proceeds. At bare minimum, there is a fact of whether the fight takes place or not. If both you and I are right—you say there is no truth-value, I say there is—we should abandon the objectivity of truth. We could only speak of 'truth-for-me' and 'truth-for-you' (Putnam, 1967, p. 244). Physical events like fights are very real phenomenon: it cannot be merely relative to our utterances whether they take place or not.

The latest defense of the Rietdijk-Putnam argument comes from Daniel Peterson and Michael David Silberstein (2010). They are careful to separate two different notions: the 'reality value' and the 'reality relation'. The former represents the ontological status of an event. It has either the value of one (real) or zero (irreal). The latter is used to define co-reality. The reality relation exists between events that have the same reality value. Events have unique reality values so an event cannot be both existent and non-existent; that would be a contradiction. The reality value and the reality relation clarify the definitions of presentism and eternalism. Under presentism, the present has the reality value of one, whereas the past and the future have reality values of zero. Under eternalism, the past, the present and the future all have the same reality value of one (Peterson & Silberstein, 2010, p. 212).

'If it is possible for one to construct a hyperplane of simultaneity', Peterson and Silberstein (2010, p. 213) note conditionally, 'between any two or more events, then these events are said to be simultaneous'. When the two events have the same reality value, they are equally real. In Minkowski geometry, the invariant spacetime interval is $s^2 = c^2 (\Delta t)^2 - (\Delta x)^2$. Hyperplanes may be assigned only to spacelike regions in which $s^2 < 0$, not to lightlike regions in which $s^2 = 0$, or to timelike regions in which $s^2 > 0$. Accordingly, one cannot construct a hyperplane among lightlike or timelike separated events.

---

[1] In the original article, Putnam (1967, p. 244) talks about 'my future' and 'your past'. These denote regions above my and below your 'nows'. This is quite confusing as we could not locate the same event between my and your 'nows'. In any case, the space-fight is one unique event.

For the presentist, there is a special moment 'now' which is simultaneous with everything that exists in the entire universe. Moments preceding the 'now' do not exist anymore, and moments coming after the 'now' do not yet exist. For the eternalist, the 'now' is not ontologically privileged but as real as the past and the future. Consider three events: (1) breakfast, (2) lunch and (3) dinner. If 2 occurs 'now', the presentist thinks it is exclusively real. Having lunch is absolutely simultaneous with everything that occurs anywhere in the world. 1 is no more and 3 is not yet. If 2 occurs 'now', the eternalist thinks it is equally real with 1 and 3. What occurs 'now' is perspectival for the eternalist. We might be having lunch right 'now', but our past breakfast or future dinner exist as well. One reason for this is that there is no unique simultaneity relation among events.

Peterson and Silberstein lean on the relativity of simultaneity to support eternalism over presentism. Figure 3 depicts four events. A is 'Andy hits his toe' and B 'Betty hits her toe'. A′ denotes 'Andy shouts in pain' and B′ 'Betty shouts in pain'. In their common frame, A and B are simultaneous at time $t_1$, and A′ and B′ at time $t_2$. Hence we have

ArB

A′rB′

in which $r$ means to 'share a reality value with', or 'is equally real with'. Co-real events have the same reality value of one. Say two spaceships fly above Andy and Betty. They move to opposite directions with (humanly speaking) considerable constant velocities. In the frame of the first spaceship, marked with primed axes, Betty's hitting her toe is simultaneous with Andy's shouting in pain. In the frame of the second spaceship, marked with double primed axes, Andy's hitting his toe is simultaneous with Betty's shouting in pain. When we consider three inertial frames, the one shared by Andy and Betty, and the two different spaceship frames, we get

ArB

A′rB′

BrA′

ArB′

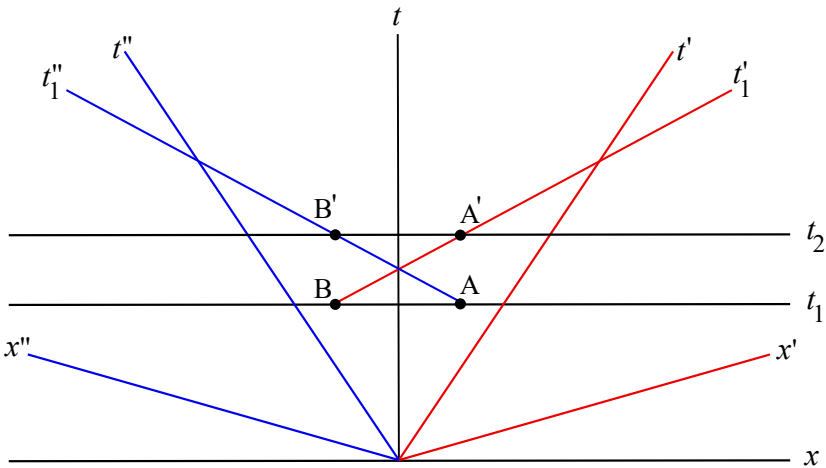

**FIGURE 3** Peterson's and Silberstein's (2010, p. 214) relativity of simultaneity spacetime diagram slightly modified

Peterson and Silberstein (2010, p. 215) note that co-reality is transitive across frames. If α*r*β, and β*r*γ, then α*r*γ. Thus,

A*r*A′
B*r*B′

An earlier event is equally real with a later event; a later event is equally real with an earlier event. All events, irrespective of their spacetime location, exist. The earlier-than and later-than relations are fixed, but locating an event in the past, in the present or in the future is relative to a frame of reference. Consequently, all tensed locations exist.

The conclusion of this argument might be criticized because we are using, perhaps in a circular fashion, the notion of 'real'. A working physicist's objection might be something like this.[2] Say an astronomer is making trajectory corrections for a satellite orbiting Jupiter. To that end, they need to know where the satellite is right 'now', how long the signal will take to reach it, and where it will be when the signal gets to it. Of course, there is no ambiguity about the 'realness' of the object at any given moment. What is 'now' to me is irrelevant for the satellite's existence. All observers possess unproblematic notion of 'realness' associated with their own spacetime diagrams. We may realize the doctrine of eternalism only after attempting to merge every disparate personal spacetime diagram into one universal spacetime. In response to such worry, it might be pointed out that the astronomer on Earth does not know where the satellite is based on stretching their own present moment throughout space. Instead, the position of the satellite is known based on its previous locations and its predicted location. There is an inference which is based on known variables at the different times the object moves along a measured/predicted trajectory. This conclusion is not reached based on the putative universality of the 'now'. There is no need to think that we should merge individual spacetime diagrams into one universal spacetime.

Carlo Rovelli (2019) sides neither with presentism nor eternalism. In his view, there is no place for a preferred, observer-independent present moment in a relativistic spacetime. The idea of a universal 'now' is a fiction, an illegitimate extrapolation of our own local experience. Rovelli claims that we should not opt for eternalism, either. He thinks the Rietdijk/Putnam, or Peterson/Silberstein type of arguments are based on a confusion of what simultaneity relations are. In the view of Rovelli, these arguments conflate obsolete Newtonian and relativistic concepts. Thus, he puts it as follows:

> Einstein's simultaneity is not a discovery of a fact of the matter about multiple simultaneity surfaces: it is the discovery that simultaneity has no ontological meaning beyond convention. (Rovelli, 2019, p. 1328)

Dennis Dieks (2012, pp. 618–619) raises the same worry. If we omit 'the well-known debate about the *conventionality* of relativistic simultaneity', which perhaps indicates that 'simultaneity is purely conventional and lacks metaphysical significance', there is 'no reason to suppose that simultaneous events share a special "reality-property"'. This means there is no reason to think simultaneous events share the same reality value of one, are equally real with each other. In the same vein, Hanoch Ben-Yami (2015) has argued that such arguments employ false premises. The reliance on hyperplanes, the critics insist, becomes a non-starter for eternalism.

---

[2] I thank the reviewer of the journal for raising this objection.

# 3 | CRITICISM CONCERNING HYPERPLANES: THE DEBATE CONCERNING THE CONVENTIONALITY OF SIMULTANEITY

We may start our analysis from Albert Einstein's (1905) paper 'On the Electrodynamics of Moving Bodies'. In the section 'Definition of Simultaneity', Einstein proposes that every judgment involving time is always a judgment about simultaneous events. He mentions an unproblematic case: the time a train arrives at a station. The arrival time is simultaneous with a specific reading of a clock. If the clock is at the same place with the train, or close to it, like at the railway platform, the question 'When does the train arrive?' does not present a difficulty. A definition like this is apt provided that 'time has to be defined exclusively for the place at which the clock is located' (Einstein, 1905, p. 142). The definition is nevertheless insufficient for cases in which the relevant events occur at different locations, especially if they are very far from each other.

This brings us to the problem of synchronization. Say we could construct two ideal clocks with the exact same constitution separated by a distance $AB$. According to Einstein's proposal, we may send a ray of light from location $A$ to $B$, which is then reflected from $B$ to $A$. The time the signal leaves from $A$ is $t_A$, which is measured by an ideal clock at $A$. The time it gets to $B$ and bounces back is $t_B$, and that time is measured by an identical clock at $B$. The time of arrival at $A$ is $t_{A'}$ measured by the clock at $A$. The two clocks are in synchrony if $t_B - t_A = t_{A'} - t_B$.

Einstein (1905, p. 143) writes that 'based on experience, we also postulate that the quantity' $\frac{2AB}{t_{A'} - t_A}$ 'is a universal constant (the velocity of light in empty space)'. The average speed of light in a closed curve is invariant: this has been empirically confirmed. The famous Michelson–Morley experiment, for instance, utilized mirrors and recorded the roundtrip time of light with the aid of an interferometer. However, as Henri Poincaré already pointed out before Einstein, there is no experimental way of ascertaining the constancy of the one-way speed.[3]

In his *Philosophy of Space and Time* (1958), Hans Reichenbach went on to argue that simultaneity of two distant events is indeterminate.[4] Say $e_A$ is the event 'Light leaves from $A$ toward $B$'. $e_B$ is the event 'Light reaches $B$ and reflects to $A$'. $e_{A'}$ is the event 'Light gets backs to $A$'. Under Reichenbach's causal theory of time, the successive temporal is fixed. A rebounding light signal establishes a causal chain. In other words, the light first leaves point $A$, second it rebounds on $B$, and third it returns to $A$. These timelike separated events occur in a definite order. Not so for spacelike separated events. Adopting Einstein's notation but adding Reichenbach's synchronization parameter $\varepsilon$, the definition of synchrony becomes $t_B = t_A + \varepsilon \left(t_{A'} - t_A\right)$, $0 < \varepsilon < 1$. If the speed of light is isotropic, $\varepsilon = \frac{1}{2}$. However, because the one-way speed of light is a postulate based on a definition, not on any brute fact about nature, the choice of the synchronization parameter is conventional (Jammer, 2006, pp. 176–178).

Adolf Grünbaum (1973, pp. 348–349) explicates the same point as Reichenbach. He produces a diagram which I have modified a little bit (Figure 4). There are six events, $E_1, E_2, E_3$, and $E_\alpha, E_\beta, E_\gamma$. $E_1$ denotes an event in which a light signal leaves from $P_1$, $E_2$ denotes an event in which the light rebounds from $P_2$, and $E_3$ denotes the event of the light getting back to $P_1$.

In the figure above, with $\varepsilon = \frac{1}{2} E_2$ is simultaneous with $E_\beta$, and with $\varepsilon = \frac{1}{4} E_2$ is simultaneous with $E_\alpha$, and with $\varepsilon = \frac{3}{4} E_2$ simultaneous with $E_\gamma$.[5] 'It is therefore', Grünbaum (1973, p. 350) encapsulates the conventionalist thesis,

---

[3]The unprovable assumption is "that light has a constant velocity, and in particular that its velocity is the same in all directions. That is a postulate without which no measurement of this velocity could be attempted. This postulate could never be verified directly by experiment" (Poincaré, 1898, p. 220).

[4]Simultaneity at the same place is unproblematic. If two things happen at the same place, the question is not about a temporal simultaneity relation but about a temporal identity relation (Reichenbach, 1958, p. 124).

[5]Say that with $\varepsilon = \frac{1}{2}$ it takes 1 s for light to reach $P_2$, and then another second to get back to $P_1$. With $\varepsilon = \frac{1}{4}$ it would then take 0.5 s for light to reach $P_2$, and then 1.5 s to get back to $P_1$. With $\varepsilon = \frac{3}{4}$ it would take 1.5 s for light to reach $P_2$, and 0.5 s to get back to $P_1$. The roundtrip $P_1 - P_2 - P_1$ with all the allowed synchronization parameters is 2 s, but the one-way trips between $P_1 - P_2$ and $P_2 - P_1$ with the synchronization parameters used in the example vary from 0.5 to 1.5 s.

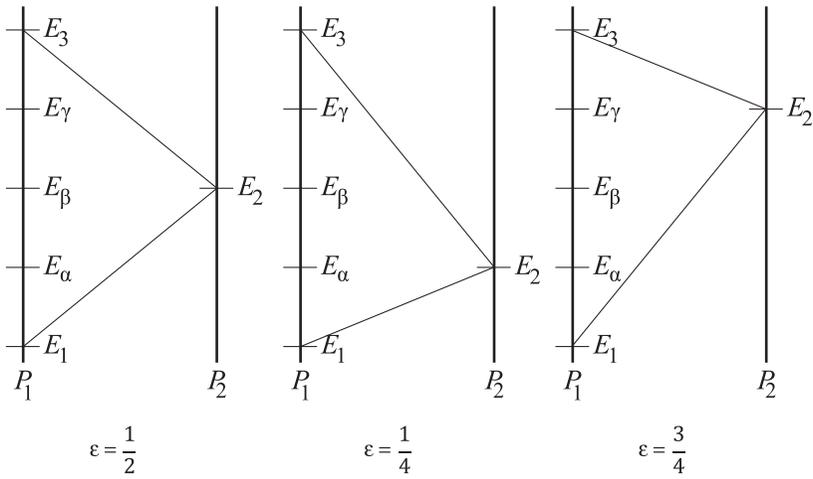

**FIGURE 4** Which events are simultaneous is conventional [Colour figure can be viewed at wileyonlinelibrary.com]

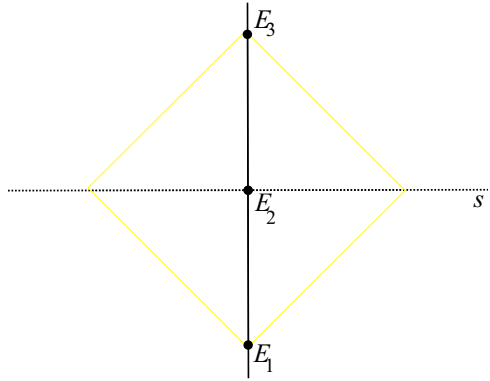

**FIGURE 5** A modification of Malament's (1977, p. 295) diagram (See Robert Trueman's philosophy of physics lecture: http://www.rtrueman.com/uploads/7/0/3/2/70324387/phil_physics_5.pdf. Date of consultation: February 14, 2022.) [Colour figure can be viewed at wileyonlinelibrary.com]

only by *convention* or definition that some *one* of these events comes to be metrically simultaneous with $E_2$, the remainder of them in the open interval $E_1 \overset{!!}{E_3}$ at $P_1$ then becoming *conventionally* either earlier or later than $E_2$.

David Malament's theorem (1977) is developed against Grünbaum's thesis, defending the non-conventionality of simultaneity within a given inertial frame of reference. In the formulation of Malament, *O* depicts an inertial observer moving through spacetime. The possible light paths from $E_1$ intersect the hyperplane *s*. If the light paths are reflected toward *O* when intersecting *s*, they will arrive on *O* at event $E_3$ (Figure 5).

Provided that the speed of light is the same in all directions, the hyperplane *s* is orthogonal to *O*. If the one-way speed of light is inconstant, the symmetry would break down and the light would reach a different event on *O* (say $E'_3$). John D. Norton (1992, p. 194) sides with Malament, taking him to have shown 'that the only nontrivial simultaneity relation definable in terms of the causal relations of special relativity is the familiar standard simultaneity relation of $\varepsilon = \frac{1}{2}$'. The debate has not so far been settled (Janis, 2018). For example, Ben-Yami (2006), Grünbaum (2010) and Robert Rynasiewicz (2012) have both provided detailed responses to the non-conventionalist thesis, supporting conventionalism.

## 4 | ETERNALISM MUST NOT HINGE ON HYPERPLANES

This section considers ways in which eternalism can be defended without an explicit endorsement of hyperplanes. The next subsection clarifies how the transitivity of reality is comprehensible in virtue of the relativity of simultaneity. The conventionality of simultaneity does not pose a problem in this context, because the relativity and the conventionality of simultaneity are altogether different notions. The subsequent subsection argues that the one-way speed of electromagnetic signals is not merely conventional because it is needed to make a difference between simultaneity and apparent simultaneity. Even though conventionality cannot be dismissed, the relativity of spacelike separation is established well enough for eternalism. And this is bad enough for presentism.

### 4.1 | Relativity ≠ conventionality

Since the introduction of the special theory of relativity, the relation between the relativity and the conventionality of simultaneity has been much disputed. The former suggests that spacelike separated events are in a definite order within a specific frame of reference. The latter suggests that spacelike separated events are not genuinely temporally ordered, not even in a frame-relative way. Some claim that the relativity and the conventionality of simultaneity are the two sides of the same coin. Vesselin Petkov (2008, p. 178) argues that

> the frame-dependency of simultaneity demonstrates that no class of absolutely simultaneous events exists. This means that we are indeed free to choose which events to regard as simultaneous since no class of events is objectively or absolutely privileged. Therefore relativity of simultaneity implies conventionality of simultaneity. The opposite is also true—conventionality of simultaneity implies relativity of simultaneity. As distant simultaneity is conventional it follows that no class of events is objectively privileged as being simultaneous; if such a class of privileged simultaneous events existed, then simultaneity could not be conventional. But as no class of events is absolutely (objectively) simultaneous, different observers in relative motion are not forced (due to the lack of a class of objectively privileged simultaneous events) to share the same class of simultaneous events, which means that simultaneity is not absolute and is therefore relative.

Petkov assimilates the conventionality thesis to the process of designating a specific inertial frame. If frames of reference are indexical notions, and there is no privileged frame, then the choice of the frame is a matter of convention. For its part, the conventionality of simultaneity, as expressed by Einstein, Reichenbach and Grünbaum, indicates that even in a specifically designated inertial frame, there is no genuine temporal order of distant events. According to conventionality, temporal order of spacelike events is indeterminate. There is no physical basis for distant simultaneity: it is up to our whim what $\varepsilon$ exactly is, provided that it is $0 >$ and $< 1$.[6]

---

[6] Some, like Gu (2021, p. 11236), point out that $\varepsilon$ could be 0 so to render the speed of light infinite in one direction, or 1 so to render it infinite in the opposite direction. Provided that the time the signal leaves from $A$ toward $B$ is 0, if $\varepsilon = 0$ in $t_B = t_A + \varepsilon(t_{A'} - t_A)$, then $t_B = t_A$. This means the one-way trip $A - B$ is done instantaneously. If $\varepsilon = 1$, then $t_B = t_{A'}$. That means the one-way trip $B - A$ is done instantaneously.

 

Based on the conventionality but not on the relativity of simultaneity, some have questioned the very existence of distant simultaneity. Long after the publication of his two relativity theories, Einstein (1949, p. 60) famously went as far as to claim that 'there is no such thing as simultaneity of distant events'. The conventionality thesis is hence radical: distant simultaneity does not exist. It is a meaningless concept.[7] The relativity of simultaneity is different because it does not destroy the temporal order of spacelike separated events. Instead, it indicates that the relation of simultaneous-with is not objective or privileged. Different observers assign different spacelike temporal orders, and no order is privileged over any other order. The relativity and the conventionality of simultaneity do not imply each other.

We should be careful as to what the relativity of a given quantity implies. Relativity is not to be equated with unreality. To show this, Slavov (2020) compares the relativity of temporal order to the relativity of electromagnetic spectrum frequency. From Lorentz transformations for time dilation, we get that if in some frame two events are simultaneous, $\Delta t' = 0$. For another frame, the same events are successive, $\Delta t \neq 0$, because $\Delta t = v\left(\Delta x'\right)/c^2\sqrt{1-(\beta)^2}$, $\Delta x' \neq 0$, $v \neq 0$, so $\Delta t$ is a positive number. In some frame the events are truly simultaneous; in another frame they are truly successive. From the Doppler effect equations, we get that a stationary source emits waves with a wavelength of $\lambda_0$, but the approaching source with a wavelength of $\lambda_+ = \sqrt{(1-\beta)/(1+\beta)}\lambda_0$, and the receding source with $\lambda_- = \sqrt{(1+\beta)/(1-\beta)}\lambda_0$. In both cases, the events and the waves are the same for all observers. What is relative is the temporal order and the received frequency. This applies both in the context of special relativity (relative motion) and in the context of general relativity (different gravitational fields).

In both cases, the relevant quantities do not have an independent, substantial existence. Tenses and colors are not out there waiting to be discovered. Their existence is perspectival. Whether something occurs now and is green in color, for example, is dependent upon the designated frame of reference. The relative and perspectival nature of some quantity does not mean that the quantity is somehow fictitious. According to eternalism, claims concerning A-predicates are indexical in nature. All events are spread across spacetime, and we occupy a contingent spacetime location.[8] The truth of tensed statements depends on the contexts of the utterance; it depends on the specific location in spacetime. This is not principally different from the way we apply indexicals in mundane circumstances. In the humdrum of our lives, we routinely say things like 'I am at home', 'Our mother is going to visit us soon', 'You just had a cup coffee', and so on. As long as the viewpoint is recognized—who makes such claims and when—the truth of indexical statements is clear enough.

Although Peterson's and Silberstein's argument relies on hyperplanes, its transitivity condition does not necessarily require them. They make a general claim: 'any relativistically invariant relational property must be transitive across all reference frames' (Peterson & Silberstein, 2010, p. 223). They mention lightlike separation of events A, B and C as an example. If A is lightlike separated from B, and C is lightlike separated from B in the same direction, then A is lightlike separated from C. This deduction holds independently of how many relativistic frames are added into the equation. 'For instance', Peterson and Silberstein (2010, pp. 223–224) note,

> if event A is light-like separated from event B in direction x in a frame moving with velocity v and event B is light-like separated from event C in direction x in a frame moving with velocity u where u is not equal to v, it is still the case that event A and event C are light-like separated in a frame moving with velocity w no matter what the value of w [as long as remaining in relativistic limits]. Thus, from this simple example, one can see that a relativistic invariant quantity is transitive across inertial frames.

---

[7]Cf. Balashov (2010, p. 56):
> Statements about simultaneity and, in general, about the temporal order of spacelike separated events are meaningless in relativistic spacetime. There is no fact of the matter as to whether one of these events occurs earlier than, simultaneous with, or later than the other.

[8]The indexicality of A-predicates and the core tenets of eternalism may be preserved even in research programs reaching beyond spacetime. Le Bihan (2020) argues that string theory implies standard eternalism, and loop quantum gravity implies atemporal eternalism.

The example above concerns lightlike separation, which is certainly different from spacelike separation. What is significant is the general point about invariance and transitivity of reality. Peterson and Silberstein argue that both definiteness and distinctness are invariant. The number of events is definite. Different observers do not disagree on the number of physical events. There are no events that are exclusively relative to a specific frame. Whether an event is past, present or future is relative; the existence of the event is not. Events exist independently of observers (Slavov, 2020, pp. 1400–1401). Definite existence is therefore invariant across frames. The Minkowski spacetime metric, $ds^2 = c^2 dt^2 - dx^2 - dy^2 - dz^2$, in which $s$ denotes the spacetime interval, and $t$ the coordinate time and $x$, $y$ and $z$ the three spatial coordinates, is invariant. It shows that the spacetime separation of distinct events is invariant. No observer could confuse two distinct events. The number and distinctiveness of physical events are both frame-transitive.

## 4.2 | Simultaneity ≠ apparent simultaneity

Marco Mamone Capria (2001, p. 777) observes that

> the standard textbook approach to special relativity implies—with no explicit justification—that, although simultaneity is frame-dependent (*relativity of simultaneity*), there is just one legitimate simultaneity notion in any given frame.

What could be the justification for the one legitimate simultaneity notion? For the purposes of our everyday life, the speed of light is infinite. We use great many technologies that utilize relativistic physics, but in observing something in our quotidian environment we may neglect the finite propagation of light. As Einstein (1936, p. 358) himself puts the point in his article 'Physics and Reality': 'We are accustomed on this account to fail to differentiate between "simultaneously seen" and "simultaneously happening".' To expound on this distinction, consider an observer $O_1$ in an inertial frame located at $x_1 = 600$ m. Lasers are turned on at $x_1 = 0$ and at $x_1 = 1200$ m. $O_1$ sees the lights at the same time. In another frame, $O_2$ is located at $x_2 = 300$ m. $O_2$ sees the lights originating from $x_2 = 0$ and $x_2 = 1200$ m at the same time. In yet another frame $O_3$ is located at $x_3 = 900$ m. $O_3$ sees the lights coming from $x_3 = 0$ and $x_3 = 1200$ m at the same time. The difference between $O_1$ to $O_2$ and $O_3$ is that in $O_1$'s frame the two separate events, namely the turning on the lasers, happen simultaneously, whereas in $O_2$'s and $O_3$'s frames the turning on the lasers are only observed simultaneously. $O_1$ sees the events in the correct order; $O_2$ and $O_3$ do not.

The distinction between 'simultaneously seen' and 'simultaneously happening' is between appearance and reality. For this distinction to work, the speed of electromagnetic signals must be isotropic. Although this cannot be proven by any crucial experiment (if there even is such a thing) or deductive inference, we should nevertheless remark that the assumption is an integral part of much of physics. Special relativity is not the most fundamental theory, but it is included in more fundamental theories, including the quantum field theory and of course the general theory of relativity. Special relativity is generally accepted by the institutional academy physics community, and it is actively used in high-energy physics. It is also the basis of various technologies, including, for example, cancer treatment, nuclear energy and GPS.

It is many times claimed there are alternative, empirically equivalent interpretations of special relativity.[9] The neo-Lorentzian interpretation retains the privileged present. This view is very far from the consensus of the physics community and hence scientific practice. Retaining absolute simultaneity and the universal 'now' requires re-interpreting science on metaphysical (Tooley, 1997) or theological (Craig, 2001) grounds. Many (e.g., Balashov & Janssen, 2003; Baron, 2018; Wütrich, 2010) have criticized this approach. One questionable assumption of neo-Lorentzianism is the postulation of an extra-structure, a privileged frame with absolute simultaneity. This structure may not be observed.

---

[9]Yet historical theories, like Lorentz's ether-theory, was not completely empirically equivalent to Einstein-Minkowski theory as the former did not fit with emerging quantum physics and general relativity (for this point, see Acuña, 2014, Section 5).

William Lane Craig thinks non-observability is a problem for verificationism, but since positivism has become obsolete, scientific discourse should not rely on verifiability. Craig (2001, pp. 130–137) maintains Einstein's verificationism is most clearly present in his operational definition of simultaneity. It reduces time to readings of clocks and measurable signals traversing between them. Although reading Einstein as an anti-metaphysical positivist has been challenged,[10] I think Craig is right in pointing out that the conventionality thesis contains some verificationist strands. My point here is not to reinterpret relativity for the sake of a novel metaphysical account. To the contrary, I have considered traditional strategies for eternalism which are rooted in the special theory of relativity. Yet I think there is an element in the conventionalist thesis that is problematic because of its positivist character. This is the problem: Is it possible to find something in a scientific theory which is *only* conventional?[11]

To say that there is something *merely* conventional is to assume an all-encompassing distinction between a strictly conventional and a strictly empirical content of a scientific theory. Before W. V. O. Quine's (1951) 'Two Dogmas of Empiricism', A. J. Ayer and Rudolf Carnap, for example, argued that mathematical and logical truths are conventional, a priori tautologies and categorically different from verifiable empirical facts. If we consider Quine's holism, however, the analytic/synthetic dichotomy becomes questionable. This does not suggest there is nothing separating the conventional and the empirical. Choosing measuring units is conventional: We may describe the motion of light by using kilometers or miles, seconds or jiffys. There are some experimental results that are very clear, like the roundtrip time of light, which has been repeatedly measured to be constant. Between the clearly conventional and the clearly empirical there is a grey area: the one-way speed of light. It involves a definition that may not be strictly verified. But a post-positivist philosophy of science should not require direct sensory verification of all the theoretical terms of a theory. Instead, the theory as a whole is tested and used in various circumstances.

In addition of being an important element of a well-established theory, the isotropy postulate is required for drawing the difference between 'simultaneously seen' and 'simultaneously happening'. This is a commonsensical distinction, comparable to other ways of correcting the information we receive with our raw senses. Without the light postulate, it is hard to see how we could otherwise maintain this distinction.

## 5 | IN SUMMARY

Classical eternalist arguments rest on ontological simultaneity and the transitivity of equal reality of spacelike separated events. These arguments postulate hyperplanes of simultaneity. Their application has been criticized due to the conventionality of simultaneity. This article explored views which are not premised on hyperplanes. Conventionality does not debunk the transitivity of reality. It is consistent to say that relativistic quantities, such as the A-location of an event, is both relative and real. All events exist equally: tenses are perspectival and indexical in nature. It was also argued that, although certain aspects of conventionality remain, distant simultaneity is not meaningless or non-existent. Within a specific inertial frame, there is a difference between simultaneity and apparent simultaneity. Events occur in different order in different frames of reference. There is no absolute simultaneity and universal present moment. That would be required for presentism. Despite its challenges, eternalism remains a compelling metaphysics of time.

**ORCID**

*Matias Slavov*  https://orcid.org/0000-0001-8515-2449

---

[10] The verificationist reading is disputed by Slavov (2019).

[11] A point like this is made by Kaila (1958).